\documentclass[sigconf,screen]{acmart}

\renewcommand\footnotetextcopyrightpermission[1]{} 
\usepackage{tikz}
\usepackage{amsmath}
\usepackage{enumitem}
\usepackage{pifont}
\usepackage{multirow,makecell}
\usepackage{color}
\usepackage{listings,amsfonts}
\usepackage{caption}
\usepackage{subcaption}
\usepackage{threeparttable}
\usepackage{bbding}
\usepackage{booktabs} 
\usepackage{longtable}
\usepackage{flushend}
\usepackage[utf8]{inputenc}

\usepackage{xhfill}
\usepackage[skins]{tcolorbox}
\usepackage{varwidth}
\usepackage{graphicx,scalerel}
\usepackage{tabularx}

\AtEndPreamble{
	\usepackage{hyperref}

}

\begin{document}

\title{Demystifying LLM Supply Chain Vulnerabilities in the Wild: Distribution, Root Cause, and Real-World Impact}


\author{Shenao Wang}
\email{shenaowang@hust.edu.cn}
\orcid{0000-0003-3818-3343}
\affiliation{%
  \institution{Huazhong University of Science and Technology}
  \city{Wuhan}
  \state{Hubei}
  \country{China}
}

\author{Yanjie Zhao}
\email{yanjie_zhao@hust.edu.cn}
\orcid{0000-0001-8793-5367}
\affiliation{%
  \institution{Huazhong University of Science and Technology}
  \city{Wuhan}
  \state{Hubei}
  \country{China}
}

\author{Zhao Liu}
\email{liuzhao3@360.cn}
\orcid{0009-0002-7535-0227}
\affiliation{%
  \institution{360 AI Security Lab}
  \city{Beijing}           
  \country{China}
}

\author{Quanchen Zou}
\email{zouquanchen@360.cn}
\orcid{0009-0004-5927-9680}
\affiliation{%
  \institution{360 AI Security Lab}
  \city{Beijing}           
  \country{China}
}

\author{Haoyu Wang}
\authornote{Corresponding author.}
\email{haoyuwang@hust.edu.cn}
\orcid{0000-0003-1100-8633}
\affiliation{%
  \institution{Huazhong University of Science and Technology}
  \city{Wuhan}
  \state{Hubei}
  \country{China}
}

\renewcommand{\shortauthors}{Shenao Wang et al.}

\begin{abstract}
Large Language Models (LLMs) are rapidly transitioning from research prototypes to core components in production systems across industries such as finance and healthcare. These deployments rely on a growing ecosystem of open-source frameworks and components, collectively forming the LLM supply chain. However, the increasing complexity of this stack introduces critical security risks that remain underexplored. In this work, we present the first systematic and large-scale empirical study of vulnerabilities in the LLM supply chain, analyzing 529 real-world vulnerabilities spanning 77 widely adopted repositories across 12 lifecycle stages. Our findings reveal that the disclosed vulnerabilities are heavily concentrated in the application layer (50.3\%) and model integration layer (42.7\%) 
Among these, 18.5\% of the vulnerabilities are LLM-specific, arising from unique architectural and workflow characteristics, such as improper handling of critical resources like model files, prompt templates, and datasets (54 CVEs, 10.2\%), as well as generative output validation errors (24 CVEs, 4.5\%). These issues expose LLM systems to severe risks, including insecure deserialization, memory corruption, and prompt injection.
To understand the real-world impact, we examine 63,243 publicly exposed LLM services and find that 45.6\% are affected by at least one remotely exploitable vulnerability, over 70\% of which are critical or high severity. By correlating these vulnerabilities with their potential exploit scenarios in the wild, we observed that these issues can lead to serious security consequences, including model tampering~(35.5\%), sensitive dataset exposure~(18.1\%), and unauthorized GPU resource abuse~(21.1\%). Based on our findings, we distill 5 actionable insights that can guide engineering teams in auditing and securing LLM services. Our work offers a data-driven foundation for securing the LLM supply chain and highlights urgent directions for both industry and future research.
\end{abstract}

\maketitle

\section{Introduction}
Large Language Models (LLMs) have ushered in a new era, redefining the capabilities of systems in natural language understanding~\cite{zhao2024llmsurvey}, text generation~\cite{roberto2023generativeai}, software engineering~\cite{hou2024llm4se,jin2024agents4se,wang2024agentsinse}, and autonomous agents~\cite{wang2024ala}. In recent years, the adoption of LLMs has rapidly transitioned from research prototypes to production-scale deployments across various industries. This shift is fueled by the availability of scalable open-source toolchains and frameworks that make it feasible for engineering teams to integrate LLM capabilities into their systems. However, real-world deployment of LLM systems introduces substantial operational complexity, as these systems are often assembled from a heterogeneous mix of components—including vector databases, inference engines, retrieval-augmented generation (RAG) pipelines, workflow orchestration frameworks, and frontend interfaces. This layered and interdependent architecture has given rise to what is now referred to as the LLM supply chain~\cite{wang2024llmsc}. As organizations scale up their LLM adoption in security-sensitive domains, understanding the risks and vulnerabilities within this supply chain is increasingly critical.

The LLM supply chain, as defined in recent studies~\cite{wang2024llmsc,hu2024llmsc,huang2024llmsc}, encompasses the infrastructure components, third-party libraries, and other dependencies that underpin the end-to-end lifecycle of LLM systems. In practice, assembling a production-ready LLM system typically involves integrating a diverse range of open-source tools and frameworks. 
For instance, developing an LLM-powered application typically requires several foundational frameworks. Frameworks like \texttt{Ollama} facilitate LLM serving, while gateways such as \texttt{LiteLLM}~\cite{litellm} provide unified interfaces for accessing different LLMs. To enable efficient data retrieval, vector databases like \texttt{Qdrant}~\cite{qdrant} support similarity search, and RAG frameworks such as \texttt{RAGFlow}~\cite{RAGFlow} integrate external knowledge into LLM workflows. Furthermore, developing LLM applications often involves frameworks like \texttt{GPTCache}~\cite{gptcache} for caching frequently used queries and \texttt{LangChain}~\cite{langchain} for orchestrating complex workflows. While the integration of open-source components accelerates time-to-market and promotes flexibility, it also significantly increases the attack surface of LLM systems. For developers and practitioners, ensuring the security, robustness, and maintainability of the LLM supply chain remains a growing challenge.

\noindent \textbf{Research Gaps.} 
Recent security reports~\cite{raycve,ollamacve,transformercve} have uncovered lots of vulnerabilities in various LLM infrastructure, highlighting the breadth and severity of these risks.
However, most existing research has predominantly focused on LLM safety, including adversarial attacks~\cite{andy2023universal}, jailbreaks~\cite{shen2024dan,xu2024jailbreak}, and backdoor attacks~\cite{li2024backdoorllm,zhao2025surveybackdoor}, which exploit vulnerabilities in the models themselves to manipulate outputs or bypass safety mechanisms. While these studies have provided valuable insights into LLM safety, they largely overlook the security properties of the underlying software systems that are far more prevalent in production environments.
In other words, engineering teams often lack the visibility and tooling to assess the security posture of the entire LLM stack.
Despite some emerging efforts to address vulnerabilities in LLM-integrated systems~\cite{liu2024llmrce,pedro2025prompt2sql,liu2025agentfuzz}, these investigations remain fragmented, and no prior work has provided a comprehensive and lifecycle-wide analysis of vulnerabilities across the LLM supply chain. For instance, it remains unclear how vulnerabilities are distributed across different stages and what root causes lead to these security issues. Moreover, the real-world security impact of these vulnerabilities remains underexplored.
Without a systematic analysis of these aspects, developers lack the empirical evidence necessary to prioritize mitigations or redesign insecure workflows.

\noindent \textbf{Our Work.} To address these gaps, we provide the first systematic empirical study of vulnerabilities in the LLM supply chain. Specifically, we collect and analyze 529 publicly reported vulnerabilities, spanning 77 open source projects. These projects encompass 12 key lifecycle stages of the LLM ecosystem. 
Through our analysis, we find that vulnerabilities are heavily concentrated in the application layer (50.3\%) and the model layer (42.7\%). While the majority of these vulnerabilities (81.5\%) are traditional issues such as access control flaws and injection vulnerabilities, 18.5\% are unique to LLM systems, which arise from the distinctive characteristics of LLMs, such as their reliance on model files, prompt templates, and the use of generative outputs in downstream systems. We systematically investigate these LLM-specific vulnerabilities and build a detailed root cause taxonomy comprising 4 categories and 15 subsets.
Finally, we quantify the real-world impact of these vulnerabilities by analyzing 63,243 publicly deployed LLM applications affected by remotely exploitable vulnerabilities. To summarize, we make the following contributions:
\begin{itemize}[leftmargin=10pt]
    \item \textbf{Systematic Study.} We conduct the first systematic empirical study of vulnerabilities in the LLM supply chain, analyzing 529 vulnerabilities reported across 77 projects. The investigation identifies 8 key findings and provides 5 actionable insights for engineering teams to better understand and mitigate LLM supply chain vulnerabilities.
    \item \textbf{Root Cause Taxonomy.} We develop a detailed root cause taxonomy for 98 LLM-specific vulnerabilities, comprising 4 categories and 11 subsets. This taxonomy provides a structured foundation for identifying, analyzing, and mitigating LLM-specific security issues.
    \item \textbf{Impact Investigation in the Wild.} We perform an in-depth investigation of the real-world impact of vulnerabilities by examining 63,243 publicly accessible LLM services, revealing the widespread presence of n-day vulnerabilities in production environments and their significant risks to critical LLM resources.
\end{itemize}

\noindent \textbf{Key Findings.}  
Our investigation reveals several critical security trends that pose substantial risks to the deployment of LLM systems.  
First, vulnerabilities are predominantly concentrated in the application layer (50.3\%) and model layer (42.7\%). The application layer, which interfaces directly with end-users and external systems, is particularly exposed, with app and front-end frameworks accounting for 39.7\% of all vulnerabilities. Meanwhile, the model layer, encompassing processes like training, optimization, and inference, exhibits the highest average number of CVEs per project (7.3), underscoring the critical importance of securing these stages of the LLM supply chain.  
Second, among the 529 CVEs analyzed, 18.5\% (98) are LLM-specific vulnerabilities, which stem from unique LLM attack surfaces, such as model files, prompt templates, datasets, and generative outputs. Improper LLM resource handling accounts for 54 CVEs (10.2\%), with insecure deserialization during model loading being the most prevalent. Generative output validation errors are another significant category, contributing 24 CVEs (4.5\%) and involving prompt injection attacks that lead to code execution, SQL injection, and XSS. These findings highlight the emerging risks introduced by the unique characteristics of LLM systems.  
Third, our analysis of 63,243 publicly accessible LLM services reveals the widespread prevalence of n-day vulnerabilities, with 45.6\% of services affected by at least one remotely exploitable vulnerability. Over 70\% of these vulnerabilities are classified as high or critical severity, and they disproportionately affect a few high-profile projects, such as \texttt{ollama}, \texttt{open-webui}, and \texttt{dify}, which collectively account for over 90\% of the vulnerable services.  
Finally, these vulnerabilities have significant real-world consequences, affecting critical LLM resources and workflows. Specifically, 60.9\% are tied to key resources such as model files (35.5\%), sensitive datasets (18.1\%), and GPU resources (21.1\%). These risks underscore the urgent need for developers to adopt holistic security measures to mitigate the growing threat landscape in the LLM supply chain.

\noindent \textbf{Data Availability}
To promote transparency and reproducibility, the dataset and resources used in this study are now available at \url{https://figshare.com/s/3336735ef865747365ed}.
\begin{figure}[t]
    \centering
    \includegraphics[width=\linewidth]{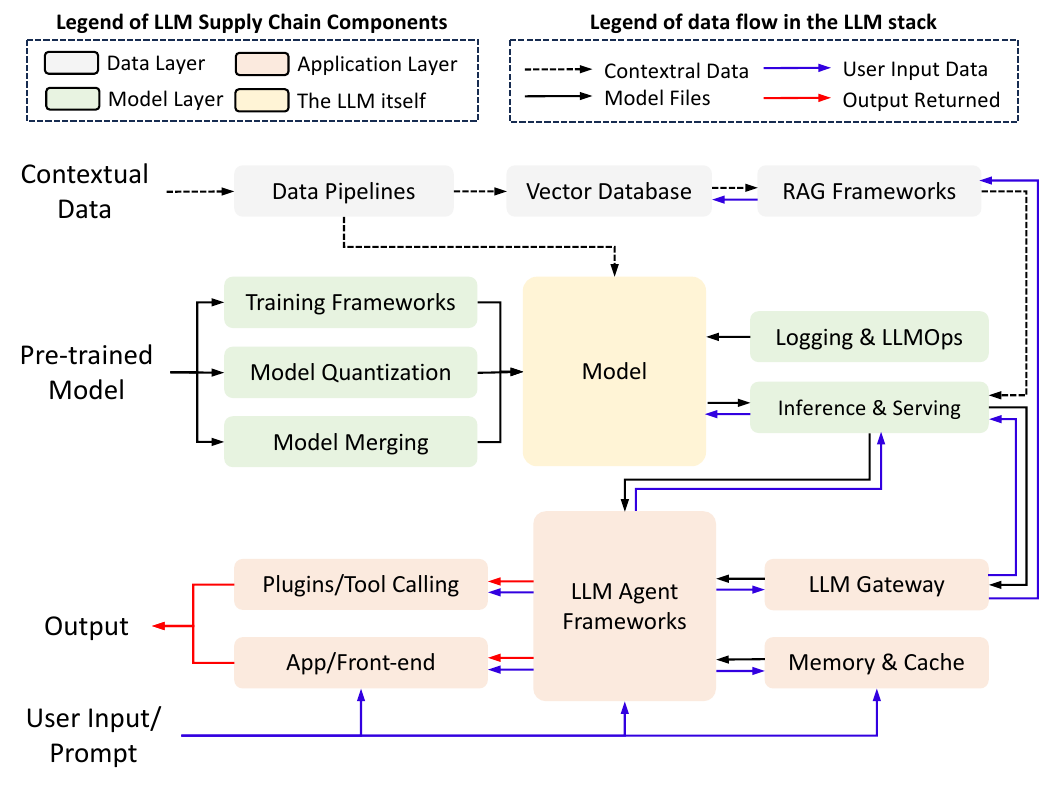}
    \caption{LLM Lifecycle and Technical Stack.}
    \label{fig:stack}
\end{figure}

\section{Background}
\label{sec:background}
Integrating LLMs into real-world systems demands a variety of technical stacks, including tools and frameworks to support the lifecycle of LLMs. Specifically, the lifecycle involves multiple interconnected stages, including data collection and preprocessing, model training, deployment, and post-deployment monitoring. As highlighted in prior studies and reports~\cite{athina-ai-tech-stack,a16z-infra-llm-app-stack}, the LLM technical stack can be broadly categorized into three layers: the data layer, the model layer, and the application layer. 

\begin{figure*}[t]
    \centering
    \includegraphics[width=0.85\linewidth]{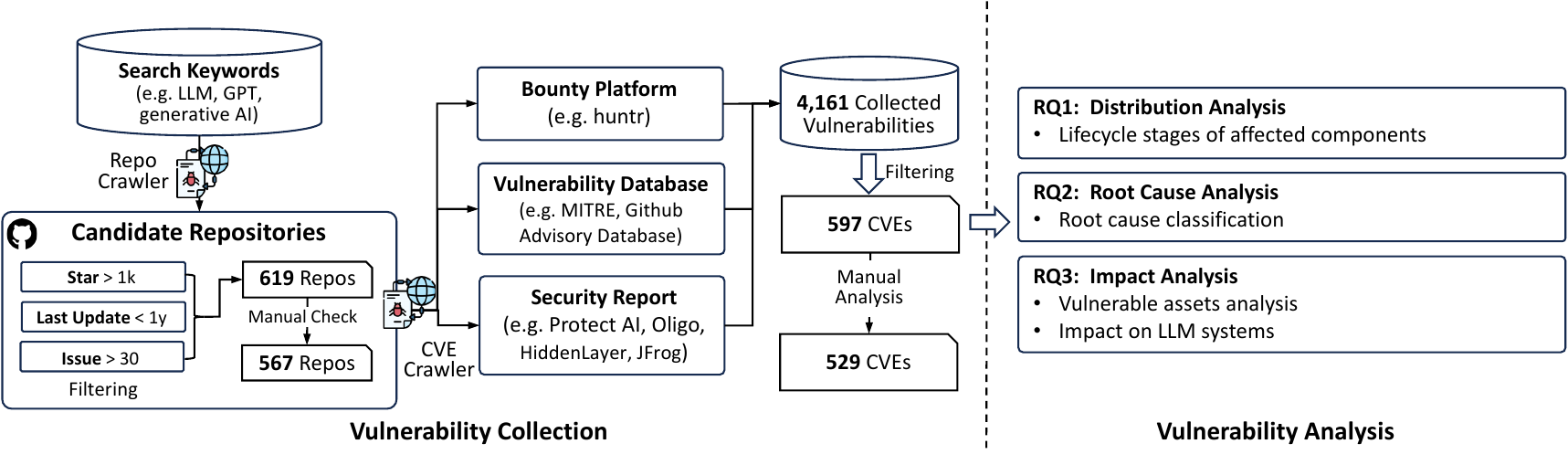}
    \caption{Overview of the methodology for analyzing security vulnerabilities in the LLM supply chain.}
    \label{fig:methodology}
\end{figure*}

\noindent \textbf{Data Layer.} As shown in \autoref{fig:stack}, The data layer serves as the foundation of the LLM supply chain, responsible for the collection, transformation, storage, and retrieval of large-scale datasets. Data ingestion involves collecting and preprocessing large-scale raw data, often using tools such as \texttt{Mage AI}~\cite{mageai} for efficient data flow automation. These data are then indexed and stored in vector databases like \texttt{FAISS}~\cite{faiss} and \texttt{Qdrant}~\cite{qdrant}, which facilitates similarity search for downstream tasks such as RAG or query caching.

\noindent \textbf{Model Layer.} The model layer involves the development, optimization, and deployment of LLMs. Frameworks like Hugging Face’s \texttt{transformers}~\cite{transformers} facilitate the implementation and fine-tuning of pre-trained models. Supporting techniques such as model quantization and model merging help optimize the model’s size and computational efficiency. LLM operations (LLMOps), such as lunary~\cite{lunary}, are also integrated into this layer, enabling continuous monitoring of the model's performance throughout its lifecycle.
Once the model is prepared, it is served and utilized through model serving and inference processes. Frameworks such as \texttt{vllm}~\cite{vllm} or \texttt{Ollama}~\cite{ollama} provide the necessary infrastructure to deploy models, enabling real-time inference via API endpoints.
    
\noindent \textbf{Application Layer.} The application layer is responsible for connecting pre-trained LLMs to real-world systems and users, enabling seamless integration and deployment. Agent frameworks like \texttt{LangChain}~\cite{langchain} and \texttt{AutoGPT}~\cite{AutoGPT} enable agent development and workflow orchestration.
Supporting tools are essential for extending the LLM’s capabilities. For example, \texttt{LiteLLM}~\cite{litellm} acts as an LLM gateway, serving as a proxy that provides a unified interface for calling multiple models in a consistent format. \texttt{GPTCache}~\cite{gptcache} provides caching services to optimize performance and reduce latency, ensuring faster responses during inference. Tools like \texttt{RAGFlow}~\cite{RAGFlow} enhance the LLM's ability to respond to complex queries by retrieving relevant information from external knowledge. Additionally, function-calling frameworks like \texttt{Composio}~\cite{composio} or MCP~\cite{mcp} can be integrated to enhance agent capabilities, allowing for dynamic interactions with external APIs and systems.
As many LLM systems interact directly with users, front-end frameworks are also a critical part of this layer. Platforms like \texttt{AnythingLLM}~\cite{anythingllm} and \texttt{LocalAI}~\cite{localai} provide interfaces for users to interact with LLMs. 
\section{Approach}
\subsection{Study Overview}
To systematically examine the prevailing security vulnerabilities in the LLM supply chain, we designed a comprehensive methodology. 
The detailed methodology is illustrated in ~\autoref{fig:methodology}.
We began by identifying and collecting repositories and artifacts relevant to LLMs and their associated components from GitHub. For each repository, we crawled vulnerabilities from established databases (e.g., MITRE CVE, GitHub Advisory Database), bug bounty platforms (e.g., \texttt{huntr}), and security reports from the community (e.g., Protect AI, Oligo, HiddenLayer). These collected vulnerabilities form the candidate dataset for manual labeling.
With the manually labeled vulnerabilities, we study 4 research questions~(RQs) focusing on distribution (RQ1), root causes (RQ2), patch effectiveness (RQ3), and real-world impact (RQ4).


\subsection{Data Collection and Pre-processing}

\noindent \textbf{Repository Identification.} Following existing works~\cite{quan2022jsdl,chen2023dlframework,shen2021compiler,chen2021faults,johirul2019bug}, we first use the GitHub search API~\cite{githubsearch} to collect repositories that are related to the LLM technical stack. We searched for repositories on GitHub using a set of keywords and their combinations, including but not limited to ``LLM'', ``pre-trained models'', ``GPT'', ``transformer'', ``agent'', and ``RAG'', etc. We included popular repositories with high star counts~(more than 1K), active maintenance~(updated within 1 year), and high developer engagement~(more than 30 issues). In total, we collected 619 candidate repositories. 
To ensure that the collected repositories accurately represent the LLM supply chain, we performed a manual check of each repository. 
Specifically, we define LLM supply chain components as the tools, frameworks, and systems that directly support the development, deployment, optimization, or operation of LLM across the 13 lifecycle stages described in \autoref{sec:background}. Repositories were included if they contributed to any aspect of the LLM lifecycle.
Furthermore, we excluded repositories such as tutorials, books, or educational materials that do not contribute directly to the LLM supply chain.
After this filtering process, we excluded 52 repositories from the initial set, resulting in a final dataset of 567 repositories.

\begin{table}[ht]
\centering
\fontsize{8}{11}\selectfont
\caption{Vulnerability data collection and processing summary. S1: Filtered out CVEs before 2022.12; S2: Filtered by relevance; S3: Deduplication; S4: Manual Labeling.}
\begin{tabular}{lrrrrrr}
\hline
\textbf{Source}              & \textbf{Raw} & \textbf{S1} & \textbf{S2} & \textbf{S3} & \textbf{S4} & \textbf{Final} \\ \hline
huntr~\cite{huntr} & 1,497              & -174        & -926        & /           & 397         & 348            \\ \hline
Github~\cite{advisory}              & 342                & -69         & -38         & -235        & 92          & 91             \\
MITRE~\cite{mitre}                        & 1,729              & -734        & -634        & -310        & 51          & 51             \\ \hline
Protect AI~\cite{protectai}                   & 392                & /           & -58         & -326        & 8           & 1              \\
JFrog~\cite{jfrog}                        & 143                & /           & -125        & -6          & 12          & 12             \\
Hidden Layer~\cite{hiddenlayer}                 & 47                 & /           & /           & -13         & 34          & 24             \\
Oligo~\cite{oligo}                        & 7                  & /           & /           & -4          & 3           & 2              \\ \hline
\textbf{Total}               & \textbf{4,157}     & \textbf{-997} & \textbf{-1,781} & \textbf{-894} & \textbf{597} & \textbf{529}   \\ \hline
\end{tabular}
\label{tab:vuln_data_summary}
\end{table}

\noindent \textbf{Vulnerability Sources}. To ensure comprehensive coverage, we collected vulnerabilities from a wide range of sources, including official vulnerability databases such as MITRE~\cite{mitre} and GitHub Advisory~\cite{advisory}, bug bounty platforms like huntr~\cite{huntr}, and security reports from the communities, such as Protect AI~\cite{protectai}. As shown in \autoref{tab:vuln_data_summary}, we crawled vulnerabilities using repository names as the basis for our search. This resulted in the collection of 1,729 vulnerabilities from MITRE and 342 from GitHub Advisory, which were further crawled to extract relevant patches and other associated information.
For \texttt{huntr}, we collected all publicly disclosed vulnerability reports, totaling 1497 vulnerabilities. Given the structured nature of these reports, we extracted key information such as vulnerability descriptions, proof-of-concept (PoC), impact, and occurrence details. Additionally, we crawled the comments of \texttt{huntr} reports, where discussions between bug hunters, platform administrators, and project maintainers often provided valuable insights into vulnerability fixes and potential discussions of similar issues, enriching our analysis.
To identify vulnerabilities directly disclosed by the community, we employed a targeted search strategy using repository names combined with keywords such as ``vulnerability'' on Google. This strategy led to the identification of four well-known companies related to the LLM supply chain security: Protect AI~\cite{protectai}, JFrog~\cite{jfrog}, Hidden Layer~\cite{hiddenlayer}, and Oligo~\cite{oligo}. We subsequently collected vulnerability reports disclosed by these organizations, which resulted in the collection of 392 reports from Protect AI, 143 from JFrog, 47 from Hidden Layer, and 11 from Oligo.

\noindent \textbf{Preprocessing}. To refine the dataset for manual analysis, we performed several preprocessing steps~(as shown in \autoref{tab:vuln_data_summary}). First, we excluded vulnerabilities reported prior to November 2022, as these were less likely to be strongly associated with LLM technologies, given that this period predates the public release of ChatGPT~(\textbf{S1}). Second, we verified whether each vulnerability was indeed associated with the 597 LLM supply chain component projects identified in our repository selection process~(\textbf{S2}). Finally, we removed duplicates across sources to avoid overrepresentation of the same vulnerability~(\textbf{S3}). After these processes, we obtained a set of 597 unique vulnerabilities for manual labeling analysis.

\subsection{Classification and Manual Labeling}
To systematically characterize vulnerabilities in the LLM supply chain, we manually labeled the 597 vulnerabilities from six aspects: (1) \textit{relevance to the LLM supply chain}, (2) \textit{affected lifecycle stage}, (3) \textit{root cause}, (4) \textit{vulnerability impact}, (5) \textit{fixing patterns}, and (6) \textit{recurrence or existence of similar vulnerabilities}. The labeling process followed an iterative approach inspired by the open coding procedure~\cite{opencoding} and prior empirical studies on software bugs and vulnerabilities~\cite{shen2021compiler,chen2023dlframework,chen2021faults,quan2022jsdl,lai2024dlvul}, ensuring both comprehensive coverage and high reliability. Below, we detail the methodology.

\noindent \textbf{Pilot Labeling.} We began with a pilot labeling phase, during which the first two authors, who have three and five years of experience in security research respectively, independently labeled a randomly selected subset (10\%) of vulnerabilities. Specifically, they follow the procedures described below. The two authors carefully read all vulnerability reports and analyzed all available information, including titles, detailed descriptions, PoCs, impact statements, associated fix patches, and any developer discussions. For each vulnerability, they assigned short but descriptive phrases as initial labels to characterize its root cause and fix strategy. After reviewing the subset of vulnerabilities, the two authors independently constructed taxonomies for root causes and fix strategies. In cases where the two authors disagreed on the categorization of a vulnerability, the third author, who served as an arbitrator, intervened to facilitate discussions and resolve conflicts. 

\noindent \textbf{Labeling Consistency Evaluation.} To ensure the reliability and consistency of the labeling process, we measured the inter-rater agreement between the two authors using Cohen’s Kappa coefficient~($\kappa$)~\cite{cohenkappa}, which is widely used in existing works~\cite{quan2022jsdl,chen2023dlframework,chen2021faults,liu2023distributed}. During the pilot labeling phase, the initial $\kappa$ score was calculated to be 0.65, indicating substantial agreement but leaving room for improvement. To address this, we conducted a training session to clarify the labeling guidelines, refine the taxonomy definitions, and resolve ambiguities in the classification process. Following the training session, the two authors independently labeled an additional 10\% of the dataset to re-evaluate the agreement. This time, the $\kappa$ score increased to 0.91, reflecting near-perfect agreement. Encouraged by this improvement, we proceeded to the full dataset labeling phase, maintaining the same process of independent labeling followed by conflict resolution. By the completion of the full labeling task, the $\kappa$ score consistently exceeded 0.8, demonstrating excellent agreement between the two authors. For vulnerabilities where disagreements arose, the authors revisited the original reports and discussed their interpretations with the third author until a consensus was reached. In summary, among the 597 manually analyzed vulnerabilities, we identified 529 that meet the criteria~(\textbf{S4}), as they are indeed related to the LLM supply chain and contain sufficient detailed information for analysis.

\noindent \textbf{Threats to Validity.}
Despite our efforts to ensure a rigorous and comprehensive data collection, preprocessing, and labeling process, several limitations remain. First, the reliance on publicly available vulnerability databases and reports, such as MITRE, GitHub Advisory, and community disclosures, might result in an incomplete dataset, as some vulnerabilities may not have been disclosed due to proprietary constraints, confidentiality agreements, or insufficient reporting by maintainers and vendors. Second, while we applied a systematic approach to filter and verify vulnerabilities, the manual nature of the relevance-checking process introduces a potential risk of human error or subjective bias. Third, our methodology for identifying vulnerabilities relies on the assumption that repository names and associated keywords are accurately indicative of their technical stack and lifecycle stage. However, this assumption may not always hold, as repositories often have limited documentation or ambiguous project descriptions, which could lead to either the inclusion of irrelevant vulnerabilities or the exclusion of relevant ones. Finally, our focus on CVEs and related vulnerability data restricts our analysis to known issues, potentially overlooking novel or emerging threats that have yet to be reported or documented. 

\subsection{Research Questions}
To better understand the security vulnerabilities in the LLM supply chain and provide actionable insights for improving the robustness of these systems, we define the following RQs:

\textbf{RQ1~(Distribution):} \textit{What are the lifecycle distributions of vulnerabilities in the LLM supply chain?} 

\textbf{RQ2~(Root Cause):} \textit{What are the common root causes of vulnerabilities in the LLM supply chain?} 


\textbf{RQ3~(Impact):} \textit{How do LLM vulnerabilities manifest in the real world and what is their potential impact in the wild?}


\section{RQ1: Distribution Analysis}
To understand the lifecycle distribution of vulnerabilities in the LLM supply chain, we analyzed 529 CVEs collected from diverse components across the lifecycle. As shown in \autoref{tab:lifecycle_components_desc}, we provide a detailed breakdown of these results.

\begin{table}[t]
\centering
\caption{Distribution of vulnerabilities across LLM supply chain components and layers.}
\label{tab:lifecycle_components_desc}
\begin{tabular}{llrrr}
\toprule
\textbf{Layer}         & \textbf{Component}               & \textbf{\#Proj}  & \textbf{\#CVE} & \textbf{\#Avg} \\ 
\midrule
\multirow{3}{*}{\textbf{Data}} 
                        & RAG Frameworks     &  2                          & 21 & 10.5  \\
                       & Vector Database      &  3                & 8 & 2.7         \\ 
                       & Data Pipeline    &  4                        & 9 & 2.3        \\ 
\midrule
\multirow{5}{*}{\textbf{Model}} 
                       & Logging \& LLMOps      & 11               & 124 & 11.3      \\ 
                       & Training Framework      &  9             & 67 & 7.4       \\ 
                      & Model Quantization     &  1              & 7 & 7         \\ 
                       & Model Inference      &  1                & 7 & 7         \\ 
                       & Model Serving         &   9              & 21 & 2.3        \\ 
\midrule
\multirow{4}{*}{\textbf{Application}} 
                       & App/Front-end       &  21                 & 210 & 10      \\
                      & LLM Gateway     &  1                     & 10 & 10        \\ 
                       & Agent Frameworks    &  12          & 41 & 3.4       \\    
                       & Plugins/Tool Calling   &   3           & 4 & 1.3         \\ 
\midrule
\textbf{Total}         &     /   &     \textbf{77}                              & \textbf{529}   &  \textbf{6.9} \\ 
\bottomrule
\end{tabular}
\end{table}

\noindent \textbf{Data Layer: 38~(7.2\%) CVEs across 9~(11\%) projects.}
The data layer, responsible for managing and processing data, accounted for 38 CVEs across 9 projects, with an average of 4.2 CVEs per project. Notably, RAG frameworks were the most affected within this layer, contributing 21 CVEs (4.0\%) from just two projects, indicating a high concentration of vulnerabilities in these systems. Vector databases, which store embeddings for semantic search or retrieval, exhibited 8 CVEs (1.5\%) across 3 projects, and these vulnerabilities often involved risks of memory corruption and denial of service~(DoS). Similarly, data pipelines accounted for 9 CVEs (1.7\%) across 4 projects, where flaws in data processing mechanisms could result in data corruption or workflow interruptions.

\noindent \textbf{Model Layer: 226~(42.7\%) CVEs across 31~(40.3\%) projects.} The model layer, encompassing processes such as training, optimization, inference, and serving, exhibited the highest average number of CVEs per project, an average of 7.3 CVEs per project. Logging and LLMOps frameworks were found to be particularly vulnerable, with 124 CVEs (23.4\%) spread across 11 projects, highlighting the critical role of operational tools in maintaining LLM workflows and the risks associated with their compromise. Training frameworks also contributed significantly, with 67 CVEs (12.7\%) across 9 projects, primarily affected by memory corruption and resource mismanagement vulnerabilities, which could undermine model accuracy or stability. By contrast, other components, such as model quantization and inference systems, exhibited a smaller number of CVEs, 7 each from a single project.

\noindent \textbf{Application Layer: 266~(50.3\%) CVEs across 37~(48.1\%) projects.} The application layer, which interfaces directly with end-users and external systems, was the most affected, resulting in an average of 7.2 CVEs per project. Front-end frameworks and applications were particularly vulnerable, accounting for 210 CVEs (39.7\%) across 21 projects, primarily due to their exposure to user inputs, APIs, and external interactions. These vulnerabilities highlight the importance of secure front-end design to prevent injection attacks, API misuse, or other user-driven exploits. Agent frameworks, which play a key role in automating workflows and integrating LLM functionalities into larger systems, accounted for 41 CVEs (7.8\%) across 12 projects, underscoring their growing importance and associated risks. Additional components within this layer, such as LLM gateways (10 CVEs, 1.9\%) and plugins or tool-calling frameworks (4 CVEs, 0.8\%), exhibited lower absolute numbers of vulnerabilities but remain critical due to their roles as intermediaries or extensions of LLM functionalities.

To further clarify how these vulnerabilities are distributed across individual frameworks and projects, we provide a detailed breakdown in the supplementary material~\cite{googledoc}. This includes the number of CVEs affecting each major LLM infrastructure project within its respective lifecycle stages.

\begin{tcolorbox}
    \textbf{Finding\#1.} The majority of disclosed vulnerabilities are concentrated in the application layer and the model layer, with App/Front-end frameworks and LLMOps being particularly vulnerable due to their exposure to user interactions.
\end{tcolorbox}
\section{RQ2: Root Cause Analysis}
To gain a deeper understanding of vulnerabilities impacting the LLM ecosystem, we analyzed 529 CVEs by classifying them into traditional and LLM-specific categories. Traditional vulnerabilities refer to those that are applicable to general-purpose software systems and are not directly tied to the unique characteristics of LLMs. In contrast, LLM-specific vulnerabilities arise from features unique to LLM architectures, such as their reliance on model files, prompt templates, or generative outputs. This allows us to identify vulnerabilities that are uniquely critical to securing the LLM supply chain, providing valuable insights for prioritizing mitigation efforts and strengthening the security of these emerging technologies.

\begin{table*}[t]
\centering
\caption{Classification of traditional vulnerability patterns and LLM-specific vulnerability patterns.}
\label{tab:cwe_distribution}
\resizebox{\textwidth}{!}{%
\begin{tabular}{c|llc|c}
\hline
\textbf{Type} & \textbf{Class} & \textbf{Root Cause} & \textbf{Count} & \textbf{Total (\%)} \\
\hline
\multirow{24}{*}{\textbf{Traditional}} 
& \multirow{4}{*}{Path Traversal}
& CWE-22: General Path Traversal & 40 & \multirow{4}{*}{121 / 22.9\%} \\
&  & CWE-29: Backslash Variant & 38 & \\
&  & CWE-23: Relative Path Traversal & 31 & \\
&  & Others (e.g., CWE-36, CWE-39) & 12 & \\
\cline{2-5}
\multirow{18}{*}{\textbf{(431 / 81.5\%)}} & \multirow{4}{*}{Injection}
& CWE-79: Cross-site Scripting & 31 & \multirow{4}{*}{91 / 17.2\%} \\
&  & CWE-78: OS Command Injection & 23 & \\
&  & CWE-94: Code Injection & 20 & \\
&  & Others (e.g., CWE-89, CWE-1336) & 17 & \\
\cline{2-5}
& \multirow{4}{*}{Improper Access Control}
& CWE-266: Incorrect Privilege Assignment & 28 & \multirow{4}{*}{76 / 14.4\%} \\
&  & CWE-862/863/639: Improper Authorization & 24 & \\
&  & CWE-352: CSRF & 15 & \\
&  & Others (e.g., CWE-648, CWE-640/306/620) & 9 & \\
\cline{2-5}
& \multirow{3}{*}{Externally Controlled Reference to a Resource}
& CWE-918: SSRF & 23 & \multirow{3}{*}{43 / 8.1\%} \\
&  & CWE-942: CORS & 8 & \\
&  & Others (e.g., CWE-73, CWE-601) & 12 & \\
\cline{2-5}
& \multirow{3}{*}{Incorrect Calculation or Exceptional Conditions}
& CWE-125/120/787: Range Error & 9 & \multirow{3}{*}{32 / 6.1\%} \\
&  & CWE-369: Divide By Zero & 8 & \\
&  & Others (e.g., CWE-476, CWE-674, CWE-312, CWE-366/367) & 15 & \\
\cline{2-5}
& \multirow{3}{*}{Uncontrolled Resource Consumption}
& CWE-770: Unrestricted Resource Allocation & 9 & \multirow{3}{*}{23 / 4.4\%} \\
&  & CWE-754: Exceptional Conditions & 9 & \\
&  & Others (e.g., CWE-1333) & 5 & \\
\cline{2-5}
& Others
& CWE-502/434/384/347/915, etc. & 45 & 45 / 8.5\% \\
\hline
\multirow{16}{*}{\textbf{LLM-Specific}} 
& \multirow{6}{*}{LLM Resource Handling Issues} 
& Insecure Deserialization in Model Loading & 28 & \multirow{6}{*}{54 / 10.2\%} \\
&  & SSTI in Prompt Template Loading  & 7 & \\
&  & Memory Corruption in Model Parsing & 6 & \\
&  & Improper Validation in RAG Resource Loading & 6 & \\
&  & Improper Validation in Dataset Loading  & 4 & \\
&  & Path Traversal in Model File Handing & 3 & \\
\cline{2-5}
\multirow{6}{*}{\textbf{(98 / 18.5\%)}} & \multirow{5}{*}{Generative Output Validation Error} 
& Prompt Injection Leads to Code Injection & 11 & \multirow{5}{*}{24 / 4.5\%} \\
&  & Prompt Injection Leads to SQL Injection & 5 & \\
&  & Prompt Injection Leads to File Manipulation & 4 & \\
&  & Prompt Injection Leads to XSS & 3 & \\
&  & Prompt Injection Leads to SSRF & 1 & \\
\cline{2-5}
& \multirow{2}{*}{LLMOps Workflow Corruption} 
& Improper Access Control in LLM-Integrated Workflows & 13 & \multirow{2}{*}{16 / 3.0\%} \\
& & Vulnerable Component in LLM-Integrated Workflows & 3 &  \\
\cline{2-5}
& \multirow{2}{*}{Distributed Training Issues }
& Insecure RPC Mechanisms & 2 & \multirow{2}{*}{4 / 0.8\%} \\
&  & Insecure Serialization and Communication Protocols & 2 & \\
\hline
\textbf{Total} & / & / & \textbf{529} & 100\% \\
\hline
\end{tabular}%
}
\end{table*}

\noindent \textbf{Overview of Traditional Vulnerabilities.}
As shown in \autoref{tab:cwe_distribution}, traditional vulnerabilities account for 431 CVEs (81.5\%) of the analyzed dataset. These vulnerabilities reflect well-established software weaknesses, including path traversal, injection, and improper access control. These weaknesses are prevalent across various software ecosystems and demonstrate the importance of adopting secure development practices. For example, traditional path traversal vulnerabilities (e.g., CWE-22) continue to pose significant risks by enabling attackers to access unauthorized resources. Similarly, injection vulnerabilities, such as SQL injection (CWE-89) or cross-site scripting (CWE-79), remain critical in web-based systems. While these vulnerabilities are not unique to LLMs, their impact can be amplified when exploited in LLM-integrated platforms.

\noindent \textbf{Overview of LLM-Specific Vulnerabilities.}
As shown in \autoref{tab:cwe_distribution}, LLM-specific vulnerabilities account for 98 CVEs (18.5\%) of the analyzed dataset. These vulnerabilities stem from the unique characteristics and workflows of LLM systems. Key factors include the reliance on model files, datasets, prompt templates, and generative outputs, which introduce new attack surfaces. Among these vulnerabilities, the most prevalent category is LLM resource handling issues~(54 CVEs, 10.2\%), which includes problems like insecure deserialization during model loading, improper validation in dataset or tool loading, and memory corruption in model parsing. The second most common category is generative output validation error, with 24 CVEs (4.5\%), highlighting issues such as prompt injection leading to code injection, SQL injection, or XSS. Other categories, such as LLMOps workflow corruption (16 CVEs, 3.0\%) and distributed training issues (4 CVEs, 0.8\%), represent lower-frequency but also critical risks, primarily stemming from insecure RPC mechanisms, improper serialization, and weak access control in LLM workflows.

\begin{tcolorbox}
    \textbf{Finding\#2.} Although traditional vulnerabilities dominate the dataset, LLM-specific vulnerabilities (18.3\%) highlight emerging risks, which are primarily introduced by attack surfaces unique to LLM systems, such as model files, prompt templates, datasets, and generative outputs.
\end{tcolorbox}

\noindent\textbf{LLM Resource Handling Issues (54 / 10.2\%).}
LLM resource handling issues represent the most prevalent category of LLM-specific vulnerabilities, accounting for 54 CVEs (10.2\%). These vulnerabilities arise from improper handling of LLM resources, such as model files, prompt templates, and datasets, which are integral to LLM workflows. Below, we describe key subsets of this category and highlight representative examples:

\begin{itemize}[leftmargin=10pt]
    \item \textbf{Insecure Deserialization in Model Loading (28 CVEs).}  Insecure deserialization is a critical subset of vulnerabilities within the category, accounting for 28 CVEs. This class of vulnerabilities arises when LLM systems load serialized model fileswithout proper validation, especially when loading some well-known vulnerable model file formats, such as torch, hdf5, and other pickle-based formats~\cite{zhao2024malhug,hiddenlayer2022pickle}. For instance, \texttt{CVE-2024-3568}\footnote{https://huntr.com/bounties/b3c36992-5264-4d7f-9906-a996efafba8f} is a deserialization vulnerability in the \texttt{transformers} library. Specifically, the vulnerability resides in the \texttt{load\_repo\_checkpoint} function of the \texttt{TFPreTrainedModel} class, where attackers could craft malicious serialized payloads (e.g., \texttt{.pickle} files) that are subsequently loaded during model checkpointing. Such vulnerabilities underscore the inherent risks in unsafe model file deserialization practices, particularly in dynamic workflows where users can autonomously load models from arbitrary sources.

    \item \textbf{Memory Corruption in Model Parsing (6 CVEs).}  
    Even when using ostensibly safe model formats, memory corruption vulnerabilities are still critical issues in the parsing of model files. These vulnerabilities often originate from improper handling of memory allocation, validation, or dereferencing during the parsing process. Such issues can lead to severe consequences, including denial of service (DoS) or even arbitrary code execution in certain cases. A representative example is CVE-2024-41130, a null pointer dereference vulnerability found in \texttt{gguf\_init\_from\_file} within the \texttt{ggerganov/ggml} library~\footnote{https://github.com/ggml-org/llama.cpp/security/advisories/GHSA-49q7-2jmh-92fp}. The vulnerability arises when the \texttt{gguf\_fread\_str} function fails to allocate memory for the \texttt{name.data} field (e.g., when reading from an invalid or corrupted \texttt{gguf} model file). In such cases, \texttt{name.data} remains a \texttt{nullptr}, and subsequent operations like \texttt{strcmp} do not check for null pointers before dereferencing.

    \item \textbf{Path Traversal in Model File Handling (3 CVEs).}  
    Path traversal is another critical issue in model file workflows, particularly during operations such as model extraction or exporting. These vulnerabilities often arise when archive files, such as those containing models or related resources, are processed without proper checks to ensure that extracted paths remain confined within the designated directories. While path traversals are common in traditional applications, they serve as a reminder to exercise caution when handling model files in LLM workflows, particularly in scenarios involving untrusted or public resources.

    \item \textbf{SSTI in Prompt Template Loading (7 CVEs).}  
    In addition to model file vulnerabilities, prompt templates represent another significant attack surface in LLM systems. Server-Side Template Injection (SSTI) vulnerabilities occur when template engines process untrusted input without proper sanitization, allowing attackers to inject and execute arbitrary code.  
    For example, \texttt{CVE-2024-2952}\footnote{https://nvd.nist.gov/vuln/detail/CVE-2024-2952} highlights an SSTI vulnerability in the \texttt{BerriAI/litellm} library, where the \texttt{chat\_template} parameter from a \texttt{tokenizer\_config.json} file is processed by the Jinja template engine without adequate validation, enabling malicious config files to execute arbitrary code on the server.  
    As prompt templates are integral to LLM workflows, rendering them should only use trusted inputs or be performed in a sandboxed environment to mitigate such risks.

    \item \textbf{Risks in RAG Resource and Dataset Loading.}  
    Beyond model files and prompt templates, RAG resource loading and dataset loading introduce additional attack surfaces in LLM workflows. These vulnerabilities often arise when external knowledge bases, vector stores, or datasets are integrated into LLM systems without proper validation. Malicious or corrupted resources can lead to issues such as unauthorized data access, data poisoning, or even code execution during preprocessing or runtime. 
\end{itemize}

\begin{tcolorbox}  
    \textbf{Finding\#3.}  
    LLM systems rely on external resources such as model files, prompt templates, datasets, RAG resources, and tools. To mitigate risks like malicious code injection or poisoning, these resources must come from trusted sources or be processed in sandboxed environments.
\end{tcolorbox}

\noindent\textbf{Generative Output Validation Errors (24 / 4.5\%).}  
Generative output validation errors represent a significant category of LLM-specific vulnerabilities, accounting for 24 CVEs. These vulnerabilities arise when model-generated outputs, such as code, database queries, or web content, are improperly validated before being used in downstream security-sensitive operations. By treating LLM outputs as inherently trustworthy, systems become vulnerable to various injection attacks, where adversarial prompts can manipulate outputs to execute unauthorized actions. Below, we describe key subsets of this category:

\begin{itemize}[leftmargin=10pt]
    \item \textbf{Prompt Injection Leads to Code/Command Injection~(11 CVEs).} 
    When LLM-generated outputs are used to produce executable code or commands, prompt injection can lead to code/command injection vulnerabilities. For example, \texttt{CVE-2023-39662}\footnote{https://github.com/run-llama/llama\_index/issues/7054} demonstrates how prompt injection can lead to remote code execution (RCE) in \texttt{llama\_index}. In this case, the vulnerability arises from the unsafe use of the \texttt{exec} function to execute Python code generated by the model. The lack of proper validation and sandboxing of the LLM-generated code allows attackers to inject malicious instructions through crafted prompts. When deployed as part of an application backend, such as a web app or Slackbot, this vulnerability exposes the server to remote exploitation. 

    \item \textbf{Prompt Injection Leads to SQL Injection~(5 CVEs).}  
    Prompt injection can result in SQL injection when model-generated outputs are directly integrated into database queries without sanitization. For instance, \texttt{CVE-2024-8309}\footnote{https://huntr.com/bounties/8f4ad910-7fdc-4089-8f0a-b5df5f32e7c5} demonstrates a prompt injection vulnerability in the \texttt{GraphCypherQAChain} class of the \texttt{LangChain} library. This vulnerability allows attackers to manipulate generative model outputs to inject arbitrary Cypher queries into a Neo4j database. Without proper validation, the execution of these malicious queries enables attackers to perform unauthorized actions such as data exfiltration or modification. 
    
    \item \textbf{Prompt Injection Leads to XSS~(3 CVEs).}  
    Prompt injection can also result in XSS when model-generated outputs are rendered into web pages without proper sanitization or escaping of special characters. For example, \texttt{CVE-2024-1602}\footnote{https://huntr.com/bounties/59be0d5a-f18e-4418-8f29-72320269a097} demonstrates how prompt injection can lead to XSS in the \texttt{LoLLMs-WebUI} application. If the malicious payload is processed and displayed in a web interface without escaping, the JavaScript code is executed within the victim's browser context. 
\end{itemize}

\begin{tcolorbox}
\textbf{Finding\#4.} Generative output validation errors in LLM-integrated systems arise from unvalidated model-generated outputs, such as code, queries, or web content, which attackers can manipulate through crafted prompts to execute injection attacks.
\end{tcolorbox}

\noindent\textbf{LLMOps Workflow Corruption (16 / 3.0\%).}  
Unlike the previous two categories, which represent new attack vectors introduced by LLM-specific functionalities, vulnerabilities in LLMOps workflows stem from traditional attack vectors that take on new significance when applied to LLM-integrated systems. These issues, including improper access control and insecure components, are particularly impactful in the context of LLMOps platforms, where complex workflows and multi-role environments amplify the risks. As these attack vectors now directly affect LLM workflows, we classify them as LLM-specific. Below, we outline two key subsets of this category:

\begin{itemize}[leftmargin=10pt]
    \item \textbf{Improper Access Control in LLM-Integrated Workflows (13 CVEs).}  
    Insufficient access control mechanisms are a recurring issue in LLMOps platforms, where inadequate role-based access control (RBAC) or missing protections can lead to privilege escalation and unauthorized actions. For example, \texttt{CVE-2024-5389}\footnote{https://huntr.com/bounties/3ca5309f-5615-4d5b-8043-968af220d7a2} highlights an Insecure Direct Object Reference (IDOR) vulnerability in the \texttt{lunary-ai/lunary} framework. This vulnerability allows a user from one organization to create, edit, or delete prompts in datasets belonging to other organizations. Exploiting such a vulnerability can severely compromise system integrity, enabling attackers to overwrite legitimate prompts, delete critical resources, or tamper with experiment results.  

    \item \textbf{Vulnerable Components in LLM-Integrated Workflows (3 CVEs).}  
    Insecure components within LLM workflows can expose systems to significant risks, particularly when unsafe functions are used in sensitive operations. A notable example is \texttt{CVE-2024-45846}\footnote{https://nvd.nist.gov/vuln/detail/CVE-2024-45846}, which arises from the use of an unprotected \texttt{eval} function in the MindsDB platform, which allows arbitrary code execution when processing crafted \texttt{SELECT WHERE} clauses in the Weaviate integration. This vulnerability underscores the dangers of relying on insecure components in LLM workflows.
\end{itemize}

\begin{tcolorbox}  
    \textbf{Finding\#5.}  
    While rooted in traditional attack vectors, vulnerabilities in LLMOps workflows, such as weak access control and insecure components, have a profound impact on LLM-specific systems.
\end{tcolorbox}

\noindent\textbf{Distributed Training Issues (4 / 0.8\%).}  
Distributed training environments introduce unique challenges and vulnerabilities in LLM-integrated systems, particularly when insecure serialization and communication protocols are used. These issues often arise in frameworks that rely on unsafe mechanisms to transmit data between nodes, thereby enabling attackers to exploit serialization flaws. While serialization vulnerabilities have long been a concern in distributed systems, their application to LLM workflows amplifies the risks by targeting sensitive training data, models, and infrastructure. Below, we describe two key subsets of this category:

\begin{itemize}[leftmargin=10pt]
    \item \textbf{Insecure RPC Mechanisms (2 CVEs).}  
    Remote Procedure Call (RPC) frameworks in distributed training pipelines can be exploited due to insecure mechanisms. For instance, CVE-2024-5480\footnote{https://huntr.com/bounties/0e870eeb-f924-4054-8fac-d926b1fb7259} in PyTorch's distributed RPC framework arises from insufficient verification in \texttt{torch.distributed.rpc}. Workers can send malicious \texttt{PythonUDF} objects via \texttt{rpc.rpc\_sync}, which the master node executes using \texttt{\_run\_function}. This enables attackers to invoke dangerous functions like \texttt{eval} and execute arbitrary commands. Such vulnerabilities expose distributed training environments to significant risks, including the compromise of sensitive data and infrastructure.

    \item \textbf{Insecure Serialization and Communication Protocols (2 CVEs).}  
    Serialization vulnerabilities can have far-reaching consequences in distributed training frameworks, as demonstrated by the \texttt{Ray RCE Vulnerability}\footnote{https://github.com/ray-project/ray}. In Ray version \texttt{2.9.1}, both the client and server use Python’s \texttt{pickle} module for communication on port \texttt{10001}. An attacker can exploit this by sending a malicious serialized payload to the server, enabling unauthenticated RCE. Furthermore, a compromised server can propagate the attack to all connected clients by sending back malicious payloads, compromising developer machines or other servers in the network. This highlights the cascading risks of insecure serialization in distributed training setups, where communication flaws can compromise not only individual nodes but the entire distributed system.
\end{itemize}

\begin{tcolorbox}  
    \textbf{Finding\#6.}  
    Distributed training environments in LLM workflows are particularly vulnerable to insecure RPC mechanisms and serialization protocols, which enable attackers to execute arbitrary code, compromise sensitive data, and disrupt distributed systems.
\end{tcolorbox}
\begin{table*}[t]
\centering
\caption{Top 8 projects with fingerprints, version sources, total services, alive services, associated CVEs, and SVPs.}
\label{tab:project_fingerprints}
\begin{threeparttable}
\begin{tabular}{lllrrrr}
    \hline
    \textbf{Project Name} & \textbf{Fingerprint} & \textbf{Version Source} & \textbf{Total}$^*$ & \textbf{Alive}$^*$ & \textbf{CVEs} & \textbf{SVPs}$^*$ \\
    \hline
    ollama             & \texttt{body="Ollama is running"} & \texttt{/api/version} & 100{,}706 & 8{,}103  & 17 & 5{,}534 \\
    open-webui         & \texttt{body="title="Open WebUI""} & \texttt{/api/version} & 67{,}490  & 15{,}873 & 36 & 12{,}097 \\
    dify               & \texttt{icon\_hash="97378986"} & \texttt{/console/api/version} & 41{,}438  & 16{,}077 & 17 & 8{,}809 \\
    gradio             & \texttt{body="<script>window.gradio\_config"} & HTML & 26{,}610  & 5{,}150  & 6  & 454 \\
    ragflow            & \texttt{body="<title>RAGFlow</title>"} & \texttt{/version} & 4{,}791   & 1{,}535  & 12 & 1{,}288 \\
    anythingllm        & \texttt{body="<title>AnythingLLM</title>"} & \texttt{/utils/metrics} & 4{,}570 & 1{,}344  & 49 & 48    \\
    langflow           & \texttt{body="<title>Langflow</title>"} & \texttt{/api/v1/version} & 2{,}321 & 987 & 6  & 40 \\
    tensorboard        & \texttt{body="<title>TensorBoard</title>"} & HTML & 1{,}885 & 317 & 3  & 0 \\
    \hline
    \textbf{Total}$^\dagger$ & / & -- & \textbf{353{,}211} & \textbf{63{,}243} & \textbf{233} & \textbf{28{,}870} \\
    \hline
\end{tabular}
\begin{tablenotes}
\item[$^*$]{The counts for Total, Alive, and SVPs columns are service-based rather than vulnerability-based.}
\item[$^\dagger$]{The Total row aggregates data across all 21 analyzed projects, not just the top 8 listed above.} 
\end{tablenotes}
\end{threeparttable}
\end{table*}

\section{RQ3: Vulnerability Impact Analysis}
To analyze the real-world impact of these vulnerabilities, we assessed publicly exposed  running vulnerable components and evaluated the specific security consequences. Below, we provide a detailed breakdown of these aspects.

\noindent \textbf{Vulnerable Services.} 
To quantify the exposure of vulnerable LLM components, we analyzed the relationship between vulnerabilities, the projects they affect, and their publicly accessible web services. Among the 529 vulnerabilities identified across 77 projects, we manually reviewed each case and focused our analysis on 233 remotely exploitable vulnerabilities that affect 21 projects. For these projects, we constructed web service fingerprints based on project-specific characteristics (e.g., HTTP response headers, content patterns, icon hashes), as detailed in \autoref{tab:project_fingerprints}. We then queried FOFA~\cite{fofa}, a widely used cyberspace search engine that indexes network-exposed services based on predefined fingerprint rules, to identify publicly accessible services deployed in the wild.
As shown in \autoref{tab:project_fingerprints}, the \textit{Total} column represents the number of web service endpoints detected by FOFA that match each fingerprint. The \textit{Alive} column indicates the subset of these endpoints that responded to active HTTP probing during the validation. 
To assess the extent of real-world impact, we introduced the notion of \textit{service-vulnerability pair~(SVP)}, where each pair denotes a publicly accessible web service that is affected by one specific remotely triggerable vulnerability.
We correlated the collected services with version information from vulnerability advisories to estimate whether each service is affected by specific CVEs.
The \textit{Version Source} column in \autoref{tab:project_fingerprints} indicates the sources to extract the version information for each project, typically through querying specific HTTP endpoints (e.g., \texttt{/api/version}, or embedded metadata in HTML responses).
The \textit{SVPs} column denotes the number of \textit{service-vulnerability pairs}, which is determined based on the affected versions of the vulnerabilities.

\begin{figure}[t]
    \centering
    \includegraphics[width=\linewidth]{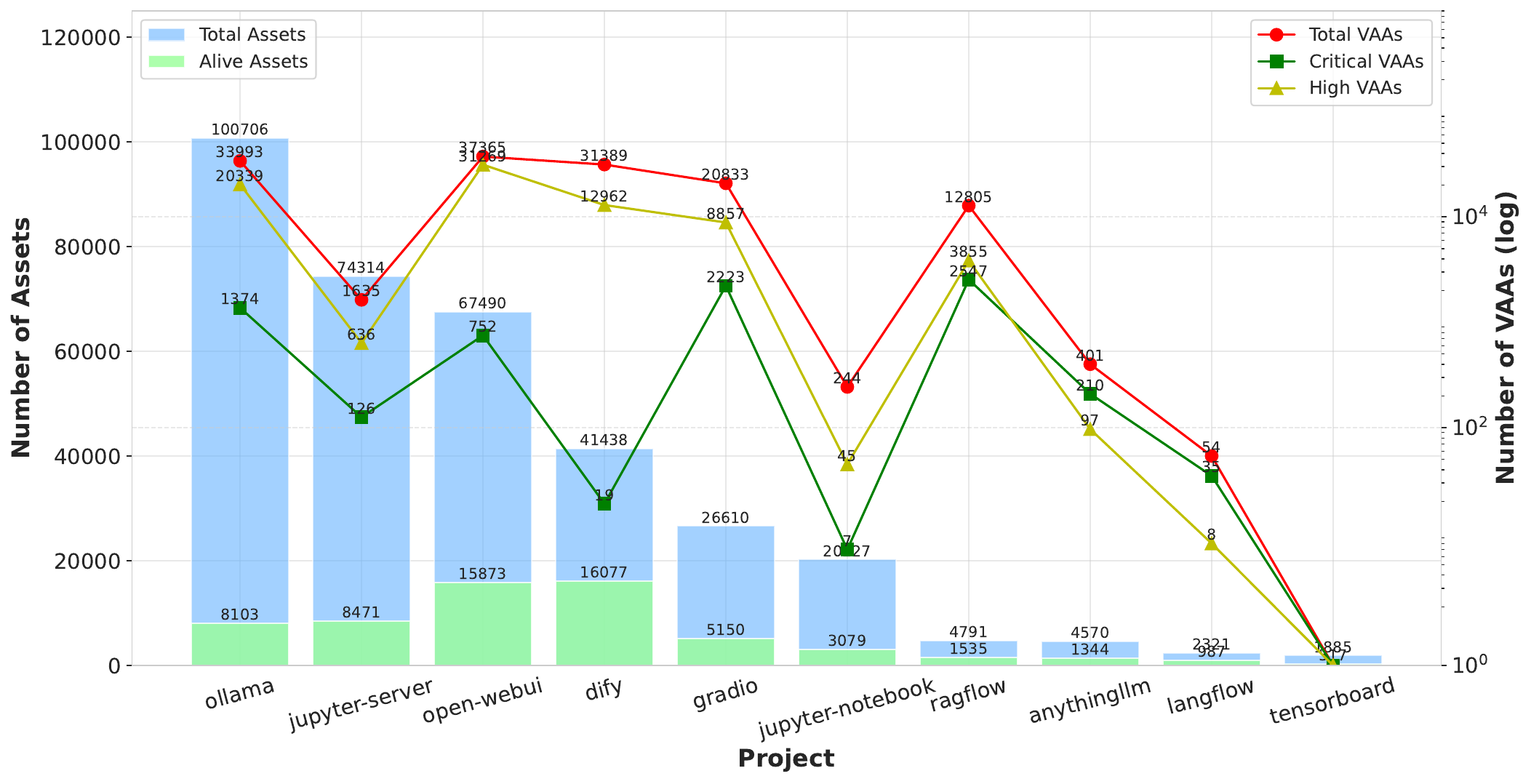}
    \caption{The SVPs across the top 8 projects. }
    \label{fig:top10_assets_vulns}
\end{figure}

\noindent\textbf{Distribution and Severity of Vulnerabilities.}  
\autoref{fig:top10_assets_vulns} illustrates the number of publicly accessible services affected by vulnerabilities across the top 8 projects. The bar chart represents the total number of publicly accessible services and the subset of alive services that are actively reachable. The line chart shows the corresponding SVPs, categorized as \textit{Low}, \textit{Medium} \textit{High}, and \textit{Critical-severity}.
Among the 353,211 identified services, 63,243 (17.9\%) are actively reachable, of which 28,870 (45.6\%) are vulnerable to at least one remotely triggerable vulnerability. Across all projects, we identified 133,411 SVPs, with 7,954~(6.0\%) are categorized as critical-severity and 86,327~(64.7\%) as high-severity, indicating that over 70\% of SVPs involve severe vulnerabilities. The exposure is disproportionately concentrated in a few high-profile projects: \texttt{ollama}, \texttt{open-webui}, and \texttt{dify} collectively account for 91.6\% of all vulnerable services. Smaller projects, such as \texttt{ragflow}, also exhibit high vulnerability rates, with 83.9\% of its 1,535 alive services being affected.

\begin{tcolorbox}  
    \textbf{Finding\#7.} Publicly accessible LLM services face significant risks, with 45.6\% vulnerable to severe CVEs and over 70\% of these impacting critical resources like model files, sensitive datasets, and GPU resources, highlighting the need for urgent security measures.
\end{tcolorbox}

\begin{table}[t]
\centering
\fontsize{8}{10}\selectfont
\caption{Impact of SVPs on LLM-integrated systems.}
\label{tab:llm_impact}
\begin{threeparttable}
\begin{tabular}{lrr}
    \hline
    \textbf{Impact} & \textbf{Affected SVPs} \\
    \hline
    Model File Integrity and Confidentiality            & 28,862                 \\
    Sensitive Dataset Integrity and Confidentiality       & 14,731  \\
    GPU Resource Availability           & 17,180            \\
    Service Availability         & 50,370                \\
    \hline
    \textbf{Total}$^*$          & \textbf{81,294}            \\
    \hline
\end{tabular}
\begin{tablenotes}
\item[$^*$]{A single SVP may lead to multiple security consequences.}
\end{tablenotes}
\end{threeparttable}
\end{table}

\noindent \textbf{Impact on LLM Systems.}  
To derive the vulnerability consequence, we first reviewed the exploit behavior for each vulnerability and then annotated the target resource types (e.g., model weights, vector indices, environment configuration, GPU resources) and attack primitives (e.g., code execution, path traversal, privilege escalation) enabled by each CVE. Using this structured metadata, we grouped vulnerabilities according to the most likely real-world consequence they introduce within an LLM service pipeline.
As shown in \autoref{tab:llm_impact}, among the 133,411 SVPs, 60.9\%~(81,294) were directly tied to critical aspects of LLM workflows. These include effects on the integrity and confidentiality of model files or datasets, the availability of the overall LLM service, and the exposure of system computing resources. 
For example, vulnerabilities like \texttt{CVE-2024-10131} and \texttt{CVE-2024-12433} in \texttt{RagFlow} collectively affected 1,165 services, allowing attackers to execute arbitrary code, which could be leveraged to tamper with pre-trained model weights or modify model inference logic. Overall, our analysis revealed that through these vulnerabilities, attackers could gain control over critical LLM resources, including model files (35.5\% of SVPs), sensitive datasets (18.1\%), and GPU resources (21.1\%). 

\begin{tcolorbox}  
    \textbf{Finding\#8.} Among the 133,411 SVPs, 60.9\% are tied to critical aspects of LLM workflows, with vulnerabilities exposing key resources such as model files (35.5\%), sensitive datasets (18.1\%), and GPU resources (21.1\%). 
\end{tcolorbox}
\section{Discussion}
\noindent \textbf{Implications.}  
Our findings highlight several critical considerations for practitioners deploying LLM-based systems in production. The first principle is to treat all externally LLM resources—including model weights, prompt templates, datasets, and configuration files—as untrusted inputs. Without rigorous validation, these artifacts can become vectors for path traversal, injection, and deserialization attacks. Second, developers must perform strict sanitization and post-generation validation of model outputs, especially in scenarios where prompts are user-influenced; prompt injection remains an underappreciated threat that can lead to downstream injection vulnerabilities. Third, when deploying LLM services (especially LLMOps services) as public-facing web applications, practitioners should not assume that default frameworks are secure. Many of them are designed for local experimentation and lack sufficient access control mechanisms, which will cause the corruption of the LLM workflow. 
Fourth, vulnerabilities in distributed training frameworks can enable attackers to compromise system integrity. Practitioners should enforce strict network segmentation in distributed environments and validate all inter-node communication to prevent unauthorized actions within the training process. 
Finally, our analysis reveals that n-day vulnerabilities remain widespread and impactful across LLM supply chains. To address these risks, organizations should implement continuous defense against n-day vulnerabilities through timely patching, while also minimizing exposure of LLM services via private deployment and strict access controls to safeguard critical resources.

\noindent \textbf{Threats to Validity.}
We acknowledge several limitations that may affect the generalizability of our findings. First, the dataset construction may introduce bias due to keyword-driven repository selection. While we used a broad set of terms, it is possible that some relevant projects were omitted. Second, our vulnerability collection method relies primarily on public databases and the use of project names as search anchors, which may miss vulnerabilities disclosed via private channels or those incompletely linked to the project name in metadata, leading to underreporting. Third, the manual labeling process, though supported by high inter-rater agreement, is inherently subjective and may be influenced by individual interpretation bias, particularly in cases where a vulnerability could plausibly belong to multiple categories. In such situations, we adopted a collaborative arbitration process to assign the most semantically appropriate classification. Lastly, the impact analysis relies on fingerprint-based asset discovery and version inference; while care was taken to construct project-specific fingerprints and version probes, the method may introduce false positives or negatives in matching. 
\section{Related Works}

\noindent \textbf{LLM Security and Privacy.} 
Recent advancements in LLMs have raised significant concerns regarding their security and privacy. 
One of the most pressing security issues in LLMs is adversarial attacks~\cite{liu2024adversary,nicholas2023align}, where attackers manipulate inputs to deceive the model into producing incorrect outputs. These attacks exploit the model’s vulnerabilities by introducing subtle, often imperceptible, perturbations to the input data~\cite{andy2023universal}, causing the model to behave unexpectedly. Additionally, jailbreaking attacks~\cite{xu2024jailbreak}, a form of adversarial attack, aim to bypass the built-in restrictions or safety mechanisms of LLMs~\cite{alexander2023jailbroken}, enabling the model to perform harmful or unintended actions~\cite{shen2024dan}. 
Backdoor attacks~\cite{zhao2025surveybackdoor} represent another critical vulnerability in LLMs. Recent studies have highlighted backdoor vulnerabilities in various contexts, such as customized GPTs~\cite{zhang2024gptsbackdoor}, RAG systems~\cite{cheng2024trojanrag}, and LLM agents~\cite{wang2024badagent}.
While existing research has primarily concentrated on content safety, our work broadens the understanding of LLM vulnerabilities by focusing on their integration within real-world systems.

\noindent \textbf{LLM Infrastructure Vulnerabilities.}
Recent studies have highlighted a range of vulnerabilities associated with the integration of LLMs into software systems. 
One prominent category of vulnerabilities is prompt injection~\cite{sahar2023promptinjection,liu2023houyi}, which can lead to severe security risks such as RCE~\cite{liu2024llmrce,liu2025agentfuzz}, SQL injections~\cite{pedro2025prompt2sql}, and even other injection vulnerabilities~\cite{icse26taintp2x}. 
These vulnerabilities arise when attackers manipulate prompts to inject malicious code or SQL queries, allowing them to exploit LLM-integrated applications. 
Additionally, pre-trained models themselves have become targets for exploitation, including PyTorch~(pickle-based)~\cite{zhao2024malhug} and TensorFlow models~\cite{zhu2025tensorflow}.
This highlights the potential for malicious actors to manipulate LLMs at the framework level, leveraging their capabilities to launch attacks. 
More recently, privacy and security risks in multi-tenant LLM environments have also been highlighted, particularly with KV-cache sharing vulnerabilities~\cite{song2024kvcache,wu2025kvcache,luo2026shadowcacheunveilingmitigating}.
Beyond the model and serving stack, recent research~\cite{wang2026vdbfuzz,xie2025understandingbugsvectordatabase,wang2025vdbms} has also begun to examine the reliability and security implications of vector databases and retrieval infrastructure used in LLM applications. 
These studies show that bugs and vulnerabilities in indexing, metadata filtering, similarity search, and access control can undermine retrieval integrity, leading to incorrect grounding results or even system crashes. 
While these studies provide important insights, our work systematically investigates vulnerabilities across the entire lifecycle, thereby offering a more comprehensive understanding.
\section{Conclusion}
In this paper, we conduct the first systematic empirical study of vulnerabilities in this emerging ecosystem. Through an analysis of 529 vulnerabilities across 77 open-source projects and 12 lifecycle stages, we show that security risks are widespread, unevenly distributed, and often rooted in LLM-specific attack surfaces.
To characterize these risks, we construct a root cause taxonomy that captures recurring patterns behind LLM-specific vulnerabilities. Furthermore, our large-scale Internet measurement of 63,243 publicly deployed LLM services demonstrates that 45.6\% of services are affected by at least one remotely exploitable vulnerability, exposing critical assets such as model files, datasets, GPU resources, and service availability.
We hope this work can serve as a foundation for future research on LLM supply chain security and help developers and practitioners build more robust and secure LLM applications.

\balance
\bibliographystyle{ACM-Reference-Format}
\bibliography{ref}

\end{document}